\documentclass[prl,twocolumn,aps]{revtex4}
\usepackage{graphicx}
\usepackage{amsmath}
\usepackage{amsfonts}
\begin{document}

\newcommand{\be}{\begin{equation}}
\newcommand{\ee}{\end{equation}}
\newcommand{\bea}{\begin{eqnarray}}
\newcommand{\eea}{\end{eqnarray}}
\newcommand{\beq}{\begin{equation}}
\newcommand{\eeq}{\end{equation}}
\newcommand{\beqn}{\begin{eqnarray}}

\newcommand{\eeqn}{\end{eqnarray}}
\newcommand{\ack}[1]{{\bf Pfft! #1}}
\newcommand{\pa}{\partial}
\newcommand{\osigma}{\overline{\sigma}}
\newcommand{\orho}{\overline{\rho}}
\newcommand{\myfig}[3]{
	\begin{figure}[ht]
	\centering
	\includegraphics[width=#2cm]{#1}\caption{#3}\label{fig:#1}
	\end{figure}
	}
\newcommand{\littlefig}[2]{
	\includegraphics[width=#2cm]{#1}}
\newcommand{\1}{{\rm 1\hspace*{-0.4ex}%
\rule{0.1ex}{1.52ex}\hspace*{0.2ex}}}

\newcommand{\nn}{\nonumber}

\title{On the 3-point functions of Aging Dynamics and the AdS/CFT Correspondence}
\author{Djordje Minic}
\affiliation{Department of Physics, Virginia Tech, Blacksburg, VA 24061, U.S.A.} 
\author{Diana Vaman}
\author{Chaolun Wu}
\affiliation{Department of Physics, University of Virginia,
Box 400714, Charlottesville, VA, 22904-4714, U.S.A}


\begin{abstract}
Aging can be realized as a sub-algebra of Schr\"odinger algebra by discarding the time-translation generator. While the 2-point functions of the Age algebra have been known for some time, little else was known about the higher $n$-point correlators. In this letter we present novel 3-point correlators of scalar primary operators. We find that the Aging correlators are distinct from the Schr\"odinger correlators by more than certain dressings with time-dependent factors, as was the case with 2-point functions. 
In the existing literature, the holographic geometry of Aging is obtained  by performing certain general coordinate transformations on the holographic dual of the Schr\"odinger theory. Consequently, the Aging 2-point functions derived from holography look as the Schr\"odinger 2-point functions dressed by time-dependent factors. However, since the 3-point functions obtained in this letter are not merely dressed Schr\"odinger correlators and instead depend on an additional time-translation breaking variable, we conclude that the most general holographic realization of Aging is yet to be found. 
We also comment on various extensions of the Schr\"odinger and Aging Algebras.
\end{abstract}

\maketitle


A remarkable example of ``non-equilibrium criticality'' is represented by
the phenomenon of aging \cite{agingbook}. 
Such ``non-equilibrium criticality''  can be observed in a feromagnetic spin system (an Ising model) prepared in a high-temperature state which after being quenched to a temperature at or below its critical temperature is left to evolve freely. It is observed that the size of the clusters of ordered spins (which form and grow) is time-dependent and scales as time to some power, the inverse of which defines the dynamical exponent. In addition, the 2-point correlation functions in such systems depend on {\it both} time values (and not {\it only} on the their difference, as it is the case in other critical phenomena which do not break time-translation) \cite{agingbook, Henkel:2010ec}. 
The essential
physics of aging (which is crucially a non-stationary process) \cite{agingbook} has been
recently discussed in the context of the AdS/CFT duality \cite{michel, Jottar:2010vp}. 
In this letter
we clearly distinguish between aging realized as
``dressed'' Schr\"odinger dynamics from pure aging. In particular we explicitly demonstrate this
difference at the level of the 3-point function.
Our {\it novel results} regarding the 3-point function should be of practical importance in both real and numerical experiments involving aging dynamics \cite{agingbook}.

Let us begin by reviewing the current understanding of holographic aging \cite{agingbook, michel, Jottar:2010vp}.
The Schr\"odinger group is the group of symmetries of the free Schr\"odinger equation
$
(2i{\cal M}\partial_t+\nabla^2)\varphi_S(t,\vec x)=0\label{sch1}.
$
The Age (or Aging) group is the same as the Schr\"odinger group, minus the time translation. To break time-translation invariance, but maintain scale invariance, 
a simple modification 
can be made to
the previous equation by adding a time-dependent potential $v(t)$ \cite{agingbook}, so that
$
(2{\cal M}[i\partial_t-v(t)]+\nabla^2)\varphi_A(t,\vec x)=0\label{age1}.
$
(For a general discussion regarding the breaking of time-translation invariance, consult \cite{agingbook}.)
However, $v(t)$ needs to transform the same way as $\partial_t$ and $\nabla^2$, so $v(t)=k/t$, where $k$ is an arbitrary constant.
The field transformation which maps these two equations into each other is given by
$
\varphi_A(t,\vec x)=\exp(-i\int^t_{t_0} d\tau v(\tau))\varphi_S(t,\vec x)=({t}/{t_0})^{-ik}\varphi_S(t,\vec x).
$
If one considers operators with a certain mass ${\cal M}$, then
$
O_A(t,\vec x)=({t}/{t_0})^{-ik{\cal M}}O_S(t,\vec x)\label{cor1}.
$
This leads straightforwardly to a relationship between the Age and Schr\"odinger $n$-point correlators:
$
G_A^{(n)}(t_i,\vec x_i;{\cal M}_i)=\prod_{i=1}^n t_i^{-ik{\cal M}_i} G_S^{(n)}(t,\vec x_i;{\cal M}_i)
$
where the $t_0$ dependence cancels due to the Bargmann selection rule $\sum_i {\cal M}_i=0$.

The holographic dual of a system which is invariant under the realization of Age algebra 
using the above trick (and which takes into account the singularity at $t=0$ in  $v(t)=k/t$) was constructed in \cite{Jottar:2010vp}. The relevant Age metric $ds_A^2$ \cite{Jottar:2010vp} reads as:
\be
ds_{A}^2=\frac{R^2}{z^2} \bigg(dz^2+\frac{2\alpha\beta}{z} dz \,dt-\frac{\beta^2}{z^2}\left(1+\frac{\alpha z^2}{\beta t}\right)dt^2-2 dt\,d\xi + d\vec x^2\bigg)\label{agemetric}.
\ee
(This geometry is locally Schr\"odinger, but its global structure is not.)
Then it is easy to check that 
$
\Phi_{A}(t,\xi,\vec x, z)=\Phi_S(t,\xi+\frac{\alpha\beta}{2}\ln\frac{\beta t}{z^2},\vec x, z)\label{age map}
$
obeys the equation of motion $\Box \Phi_A=0$ in the Age metric if $\Phi_S$ obeys the equation of motion $\Box \Phi_S=0 $ in the Schr\"odinger background
\be
ds_{S}^2=\frac{R^2}{z^2} \bigg(dz^2-\frac{\beta^2}{z^2}dt^2-2 dt\,d\xi + d\vec x^2\bigg).
\ee
(This can be extended for fields of arbitrary spin.)
In terms of the boundary field values $\bar\Phi_S$, the bulk Schr\"odinger field can be written as 
\be
\Phi_S(t,\xi,\vec x,z)=\!\!\!\int_{t',\xi', \vec x'} \!\!\!\!\!{\cal G}_S(t-t',\xi-\xi', \vec x-\vec x',z)\bar \Phi_S(t',\xi',\vec x'),
\ee
which, after applying the above map becomes
\bea
\Phi_{A}(t,\xi-\frac{\alpha\beta}2\ln\frac{\beta t}{z^2},\vec x,z)&=&\!\!\!\int_{t',\xi', \vec x'} \!\!\!\!\!{\cal G}_S(t-t',\xi-\xi', \vec x-\vec x',z) \nn \\
&&\!\!\!\!\!\!\!\times \bar \Phi_{A}(t',\xi'-\frac{\alpha\beta}{2}\ln\frac{\beta t'}{z_{b}^2},\vec x'),
\eea
where $z_b$ is the value of $z$ at the regularized boundary ($z_b\ll 1$). Here ${\cal G}_S(t-t',\xi-\xi', \vec x-\vec x',z)$ denotes the boundary-to-bulk propagator of a field in the Schr\"odinger background. 
Fourier-transforming along the $\xi$ direction, we find 
\bea
\Phi_{A}(t,\vec x,z; {\cal M})&=&\int_{t', \vec x'} \exp\left(i {\cal M}\frac{\alpha\beta}{2}\left(\ln\frac{\beta t}{z^2}-\ln\frac{\beta t'}{z_b^2}\right)\right) \nn \\
&&\times {\cal G}_S(t-t', \vec x-\vec x', z; {\cal }M)\bar \Phi_{A}(t',\vec x'; {\cal M}). \nn \\
\eea
This enables us to reconstruct, with relative ease, the holographic answer for the correlators of primary operators with respect to the Age algebra from those derived using the Schr\"odinger background holography \cite{bvwa, bvw}. For example, the 3-point function of a scalar operator is 
\be
\delta({\cal M}_1+{\cal M}_2+{\cal M}_3)\Lambda\!\!\!\!\prod_{i=1,2,3}(t_i)^{-i{\cal M}\frac{\alpha\beta}{2}}\int_{t,\vec x,z}\!\!\!{\cal G}_S(t-t_i, \vec x-\vec x_i; {\cal M}_i),\label{a 3 point}
\ee
assuming that the scalar source, $\bar\phi({\cal M})$, has a cubic coupling in the bulk: $S_{grav}=\int (\dots+  \frac{1}{3!}\Lambda \phi^3(t,\vec x, z;{\cal M})  +\dots).$ By using the Bargmann superselection rule $\sum_i {\cal M}_i=0$ (which in this holographic context is simply the momentum conservation along the $S^1$ direction parametrized by $\xi$) the 3-point function is reduced to the previous form. 
To get the Age correlator one performs a DLCQ projection along $\xi$ followed by a functional differentiation with respect to the boundary fields.
This guarantees that the Age correlators differ from the Schrodinger correlators only by time-dependent phase factors, if the tensor indices are in the $\vec x$ directions, as in the case of scalar operators.

{\it One of the main points of this letter is that this realization of aging does
not capture the most general aging dynamics and that what has been described above is
just a particular realization of the Schr\"odinger dynamics!}
In what follows we clearly distinguish between this special case and the most
general aging dynamics.

For simplicity, let us consider a 1+1-dimensional theory with coordinates $t,r$. We use $\xi$ to denote the Fourier variable conjugate to the mass ${\cal M}$ of a certain primary operator. In the notation of \cite{agingbook, hu}, the Schr\"odinger and Age algebras are respectively spanned by the generators $\{X_{-1}, X_0, X_1, M_0, Y_{\frac{1}{2}}, Y_{-\frac{1}{2}}\}$ and $\{ X_0, X_1, M_0, Y_{\frac{1}{2}}, Y_{-\frac{1}{2}}\}$, which obey the following commutation relations 
\bea
&&\!\!\!\!\![X_n, X_{n'}]=(n-n')X_{n+n'}, \;\;[X_n, Y_m]=(\tfrac {n}{2} -m)Y_{n+m}\nn\\
&&\!\!\!\!\![X_n, M_{n'}]=-n' M_{n+n'}, \;\; [Y_m, Y_{m'}]=(m-m')M_{m+m'}.\nn\\
\eea
The most general realization of these generators (which is apparently {\it new}) is
\bea
X_{-1}&=&-\partial_t+\frac{g(t)-\gamma}{t}+\frac{h(t)-\delta}{t} i \partial_\xi\nn\\
X_0&=&-t\partial_t -\tfrac 12 r \partial_r -\tfrac {\Delta}2 + g(t)+h(t) i\partial_\xi\nn\\
X_1&=&-t^2 \partial_t - tr\partial_r - \Delta t + \tfrac i2 r^2 \partial_\xi \nn\\&+& t(g(t)+\gamma)+t(h(t)+\delta)i\partial_\xi\nn\\
Y_{-\frac{1}{2}}&=&-\partial_r, Y_{\frac{1}{2}} = -t\partial_r + ir \partial_\xi,  M_0=i\partial_\xi\equiv -{\cal M},
\label{agerealiz}
\eea
where $g(t), h(t)$ are arbitrary time-dependent functions and $\gamma,\delta$ are arbitrary constants. In arriving at (\ref{agerealiz}) we have kept the form of the generators $M_0$ and of the spatial translation $Y_{-\frac{1}{2}}$ and generalized Galilean-invariance $Y_{\frac{1}{2}}$ unchanged. This general realization is central for the new results presented in what follows.

Next we solve the partial differential constraints imposed on the 3-point functions (namely that they are left invariant by the Age generators). The conclusion is that the most general scalar 3-point function is
\bea
&&\!\!\!\!\!\!G_{A}(\{t_i, r_i, \xi_i\})=\bigg(\prod_{i=1}^3 t_i^{\gamma_i}\bigg)
\exp(\sum_{i=1}^3 \int^{t_i} d\tau\frac{g(\tau)}\tau)\nn\\
&&\!\!\!\!\!\!\times
(t_3-t_1)^{-\tfrac 12 \Delta_{31,2}+\gamma_{31,2}} (t_3-t_2)^{-\tfrac 12 \Delta_{32,1}+\gamma_{32,1}}\nn \\
&&\!\!\!\!\!\!\times
(t_2-t_1)^{-\tfrac 12 \Delta_{21,3}+\gamma_{21,3}}  \Theta_A(u_1, u_2, u_3, \frac{t_3(t_2-t_1)}{t_2(t_3-t_1)})\label{mostgenage}
\eea
where 
$
\gamma_{31,2}=\gamma_3+\gamma_1 -\gamma_2 \;{\rm etc} $ and
$\Delta_{31,2}=\Delta_3+\Delta_1 -\Delta_2 \;{\rm etc}$.
Here $\Theta_A(u_1, u_2, u_3, \frac{t_3(t_2-t_1)}{t_2(t_3-t_1)})$ is some unconstrained function of:
\bea
&&u_1=-2i(\xi_2-\xi_1) + \frac{(r_2-r_1)^2}{t_2-t_1}+2\int_{t_1}^{t_2} d\tau \frac{h(\tau)}{\tau}\nn\\
&& + 2(\delta_2-\delta_1)\ln(t_2-t_1)
+2(\delta_1+\delta_2)\ln\frac{t_3-t_2}{t_3-t_1}\nn\\
&&-2\delta_2\ln t_2+2\delta_1\ln t_1,\nn\\
&& u_2=
  -2i(\xi_3-\xi_1) + \frac{(r_3-r_1)^2}{t_3-t_1}+2\int_{t_1}^{t_3} d\tau \frac{h(\tau)}{\tau}\nn\\
&& + 2(\delta_3-\delta_1)\ln(t_3-t_1)+2(\delta_1+\delta_3)\ln\frac{t_3-t_2}{t_2-t_1}\nn\\
&&-2\delta_3\ln t_3+2\delta_1\ln t_1,\nn\\
&&u_3= -2i(\xi_3-\xi_2) + \frac{(r_3-r_2)^2}{t_3-t_2}+2\int_{t_2}^{t_3} d\tau \frac{h(\tau)}{\tau}\nn\\
&& + 2(\delta_3-\delta_2)\ln(t_3-t_2)+2(\delta_2+\delta_3)\ln\frac{t_3-t_1}{t_2-t_1}\nn\\
&&-2\delta_3\ln t_3+2\delta_2\ln t_2.
\eea
For the Schr\"odinger 3-point correlator we find a similar expression, but without the dependence on the additional variable
$\frac{t_3(t_2-t_1)}{t_2(t_3-t_1)}$:
\bea
&&\!\!\!\!\!\!G_{S}(\{t_i, r_i, \xi_i\})=\bigg(\prod_{i=1}^3 t_i^{\gamma_i}\bigg)
\exp(\sum_{i=1}^3 \int^{t_i} d\tau\frac{g(\tau)}\tau)\nn\\
&&\!\!\!\!\!\!\times
(t_3-t_1)^{-\tfrac 12 \Delta_{31,2}+\gamma_{31,2}} (t_3-t_2)^{-\tfrac 12 \Delta_{32,1}+\gamma_{32,1}}\nn \\
&&\!\!\!\!\!\!\times
(t_2-t_1)^{-\tfrac 12 \Delta_{21,3}+\gamma_{21,3}}  \Theta_S(u_1, u_2, u_3).\label{mostgensch}
\eea
Note that despite the presence of the time-dependent prefactors this correlator is time-translation invariant.
In fact, it is easy to check that a redefinition of the primary fields of the Schr\"odinger algebra, effected by factoring out appropriate time-dependent functions, gives the correlators of the type (\ref{mostgensch}).
However, this redefinition does not change the fact that $X_{-1}G_S(\{t_i, r_i, \xi_i\})=0$. The fundamental difference between Age and Schr\"odinger 3-point functions lies in the dependence of the former on
$\frac{t_3(t_2-t_1)}{t_2(t_3-t_1)}$. 

At this stage we pause to note that the analysis performed at the beginning of this letter regarding the form of the Age correlators was too restrictive.
Since the time-dependent potential is introduced by a simple redefinition of the fields, $\varphi_S(t,\vec x)=\exp(i\int^t d\tau v(\tau)) \varphi_A(t, \vec x)$, the relevant symmetry group is still the full Schr\"odinger and {\it not} the Age group. One of the consequences of this observation is that the holographic realization of Aging (\ref{agemetric}) is equally restrictive, and thus, the most general holographic Age background is yet to be found. This is further evidenced by the fact that the 3-point correlators implied by the holographic Age metric (\ref{agemetric}) are dressed Schr\"odinger correlators (i.e. they are ``fake'' Age correlators), whereas the ones in (\ref{mostgenage}) are not.

For completeness, we also present the 3-point correlators of the scalar fields in terms of their masses:
\bea
&&\!\!\!\!\!\!\!
G_A(\{t_i, r_i, {\cal M}_i\})=2\pi\delta(\sum_{i=1}^3 {\cal M}_i)\bigg(\prod_{i=1}^3
t_i^{-\gamma_i +{\cal M}_i \delta_i}\bigg)\nn\\
&&\!\!\!\!\!\!\times\exp\bigg(\sum_{i=1}^3 \int^{t_i} d\tau\frac{g(\tau)-{\cal M}_ih(\tau)}\tau\bigg)\nn\\
&&\!\!\!\!\!\!\times
(t_3-t_1)^{-\tfrac 12 \Delta_{31,2}+\gamma_{31,2}-({\cal M\delta})_{31,2}}\nn\\
&&\!\!\!\!\!\!\times (t_3-t_2)^{-\tfrac 12 \Delta_{32,1}+\gamma_{32,1}-({\cal M}\delta)_{32,1}}
\nn\\
&&\!\!\!\!\!\!\times
(t_2-t_1)^{-\tfrac 12 \Delta_{21,3}+\gamma_{21,3}-({\cal M}\delta)_{21,3}} \nn \\
&&\!\!\!\!\!\!
\times \exp\bigg(-\frac{{ \cal M}_2 (r_2-r_1)^2}{2(t_2-t_1)}-\frac{{\cal M}_3(r_3-r_1)^2}{2(t_3-t_1)}\bigg)
\nn \\
&&\!\!\!\!\!\!\times
\tilde\Theta_A({\cal M}_2, {\cal M}_3,w,
\frac{t_3(t_2-t_1)}{t_2(t_3-t_1)}). \label{mostgenageM}
\eea
Here
$
({\cal M}\delta)_{31,2}={\cal M}_3\delta_3+{\cal M}_1\delta_1-{\cal M}_2\delta_2,
$
and
$
w=\frac{[(t_3-t_1)(r_2-r_1)-(t_2-t_1)(r_3-r_1)]^2}{(t_3-t_2)(t_2-t_1)(t_3-t_1)}.
$
For the Schr\"odinger 3-point correlators one obtains an expression similar to (\ref{mostgenageM}), but with an unconstrained function $\tilde\Theta_S=\tilde\Theta_S({\cal M}_2, {\cal M}_3,w)$.

We mention in passing that there are extensions of both Age and Schr\"odinger algebras which involve the addition of new generators that ensure the closure of the respective algebras. For example, provided that $\delta=0$ in (\ref{agerealiz}), we may add to both algebras
$
N=-t\partial_t +\xi\partial_\xi + g(t)-\gamma' + [h(t)+\int^t d\tau\frac{h(\tau)}\tau + \delta']i\partial_\xi \label{N}.
$
The non-zero commutators that involve $N$ are $[N,X_{\pm1}]=\mp X_{\pm 1}$, $[N, M_0]=-M_0$, 
$[N, Y_{\frac{1}{2}}]=-Y_{\frac{1}{2}}$.
In this case the scalar 3-point function of this extension of Age algebra is given by
\bea
&&\hat G_A(\{t_i, r_i, {\cal M}_i\})=2\pi\delta(\sum_{i=1}^3 {\cal M}_i)\bigg(\prod_{i=1}^3
t_i^{-\gamma_i}\bigg)\nn\\
&&\!\!\!\!\!\!\times\exp\bigg(\sum_{i=1}^3 \int^{t_i} d\tau\frac{g(\tau)-{\cal M}_i h(\tau)}\tau\bigg)
(t_3-t_1)^{-\tfrac 12 \Delta_{31,2}+\gamma_{31,2}}\nn\\
&&\!\!\!\!\!\!\times (t_3-t_2)^{-\tfrac 12 \Delta_{32,1}+\gamma_{32,1}}
(t_2-t_1)^{-\tfrac 12 \Delta_{21,3}+\gamma_{21,3}} \nn \\
&&\!\!\!\!\!\!
\times \exp\bigg(-\frac{{ \cal M}_2 (r_2-r_1)^2}{2(t_2-t_1)}-\frac{{\cal M}_3(r_3-r_1)^2}{2(t_3-t_1)}\bigg)\nn \\
&&\!\!\!\!\!\!\times
w^{-\tfrac 12 (\Delta_1+\Delta_2+\Delta_3)+\gamma_1+\gamma_2+\gamma_3+2}\nn \\
&&\!\!\!\!\!\!\times
\widetilde{\widehat\Theta_A}(w{\cal M}_2  , w{\cal M}_3,
\frac{t_3(t_2-t_1)}{t_2(t_3-t_1)}
 ).\nn \\
 \label{mostgenage1M}
\eea
For the Schrodinger algebra extended by $N$, we encounter again an expression similar to (\ref{mostgenage1M}) but with an unconstrained function $\widetilde{\widehat \Theta_S}=\widetilde{\widehat\Theta_S}(w{\cal M}_2, w{\cal M}_3)$.
Yet another closed subalgebra extension of Age is obtained by adding both $N$ and $V_+$, where
$
V_+ = -2t r\partial_t -2\xi r\partial_{\xi} -(r^2 + 2i\xi t)\partial_r -2 \Delta r+ 2r (g(t)+\gamma)
\label{Vp},
$
and where $h(t)=0, \delta=0$ in (\ref{agerealiz}), and $\delta'=0, \gamma'=\gamma$ in the expression for $N$.
On the other hand, adding both $N$ and $V_+$ to the Schr\"odinger algebra and requiring its closure \footnote{From $[X_{-1}, V_+]=-2i V_-$ it is clear that the Schrodinger algebra does not admit an extension which includes $V_+$ without adding all the other generators of the conformal algebra.}, leads to the full 1+2-dimensional conformal algebra in a space parametrized by $t,r , \xi$ coordinates. The generators take the form given by (\ref{agerealiz}), the expressions for $N$ and $V_{+}$, with $\delta=0, h(t)=0,\gamma'=\gamma,\delta'=0$, supplemented by
$V_-=-\xi\partial_r + ir\partial_t-\frac{ir}{t}(g(t)-\gamma)$ and
\be
W=-\xi^2\partial_\xi -\xi r \partial_r + \tfrac i2 r^2 \partial_t -\Delta\xi+2\gamma\xi
-\frac {ir^2 (g(t)-\gamma)}{2t}.
\ee
The familiar scalar 3-point function is dressed by time-dependent factors which originate in a particular realization of the generators allowing for non-zero $\gamma$ and $g(t)$:
\bea
&&\!\!\!\!\!G_c(\{t_i, r_i, \xi_i\})=\bigg(\prod_{i=1}^3 t_i^{\gamma_i}\bigg)
\exp(\sum_{i=1}^3 \int^{t_i} d\tau\frac{g(\tau)}\tau)\nn\\
&&\!\!\!\!\!\times X_{21}^{-\tfrac 12 \Delta_{21,3}+\gamma_{21,3}}
 X_{31}^{-\tfrac 12 \Delta_{31,2}+\gamma_{31,2}}
X_{32}^{-\tfrac 12 \Delta_{32,1}+\gamma_{32,1}}.\label{conf}
\eea
Here $X_{12}$, etc. are the Lorentz invariant intervals
$
X_{12}=-2i(t_2-t_1)(\xi_2-\xi_1)+(r_2-r_1)^2,  {\rm etc}.
$
However, we want to stress that the time-dependent factors in (\ref{conf}) once again do not signal any breaking of time-translation invariance, and can be obtained by redefining the primary operators of the type we have encountered earlier in this letter. 

Finally, we would like to comment on the holographic realization of the Age algebra in terms of metric isometries of a 1+3-dimensional space. The holographic dual space is parametrized by $x^\mu$ coordinates: $t, r ,\xi$ and the holographic coordinate $z$. The main observation is that once one identifies the Killing vectors $K=K^\mu \frac{\partial}{\partial x^\mu} $ obeying the Age algebra, one can reverse engineer the metric by solving the Killing vector equations for the components of the metric, i.e. $(g_{\rho\nu} \partial_\mu + g_{\rho\mu}\partial_\nu)K^\rho+K^\rho \partial_\rho g_{\mu\nu}=0$. It is natural to assume that $Y_{-\frac{1}{2}}$ and $M_0$ are bulk Killing vectors. If one makes the additional assumption that $Y_{\frac{1}{2}}$, given by (\ref{agerealiz}), is a bulk Killing vector then the problem becomes quite tractable. The bulk forms of the Killing vectors $X_{0}$ and $X_1$ are
$
X_0=-t\partial_t+X_0^\xi \partial_\xi -\tfrac 12 r\partial r 
+X_0^z\partial_z 
$
and
$
X_1=-t^2\partial_t+X_1^\xi\partial_\xi -tr\partial_r +X_1^z\partial_z,
$
where
$X_0^z=\frac{\partial_t (tg_{rr})}{\partial_z g_{rr}}$,
$X_1^z=\frac{\partial_t(t^2 g_{rr})}{\partial_z g_{rr}}$ and
\bea
&&\!\!\!\!\!\!\! X_0^\xi =\frac{i}{2t\partial_z g_{rr}}\bigg
(-\partial_z(g_{rr} S)-t\partial_t g_{rr} \partial_z S
+t \partial_t S\partial_z g_{rr} \nn\\
&&\;\;\;\;+ 2tT\partial_z g_{rr}  + 2tC_1\partial_z g_{rr}\bigg)\nn\\
&&\!\!\!\!\!\!\! X^\xi_1=\frac{i}{2\partial_z g_{rr}}\bigg(-z^2 \partial_z g_{rr}-2 g_{rr}\partial_z S - S\partial_z g_{rr}-t\partial_t g_{rr}\partial_z S \nn\\
&&+ t\partial_t S\partial_z g_{rr}+2tT \partial_z g_{rr} \bigg).
\eea
Here $g_{rr}=g_{rr}(t,z), S=S(t,z), T=T(t)$ and $C_1$ is an arbitrary constant. Solving the Killing vector equations corresponding to $Y_{-\frac{1}{2}}$ and $M_0$ leads to a metric which is $\xi, r$-independent. Furthermore,  solving the $Y_{\frac{1}{2}}$ Killing equations brings the metric to a form which coincides with the initial ansatz of \cite{Jottar:2010vp}:
$
ds^2 = g_{tt}(t,z) dt^2 + g_{rr}(t,z) dr^2 + g_{zz}(t,z) dz^2 
- 2i g_{rr}(t,z) dt d\xi + 2 g_{tz}(t,z) dt dz.
$
The other components of the metric are determined by the remaning Killing equations:
$g_{zz}=\frac{C_2 (\partial_z g_{rr})^2}{g_{rr}^2} $ and
\bea
&&\!\!\!\!\!\!g_{tz}=\frac{g_{rr}\partial_z S}{2t}+\frac{C_2 \partial_z g_{rr}\partial_t(t^2 g_{rr}) }{t^2 g_{rr}^2} + C_1 \partial_z g_{rr}\nn\\
&&\!\!\!\!\!\!g_{tt}= C_3 g_{rr}^2 + \frac{C_2(\partial_t(t^2 g_{rr}))^2}{t^4g_{rr}^2}\nn\\
&&+
\frac{2C_1\partial_t(t^2 g_{rr})}{t^2}+\frac{g_{rr}(2tT-S+t\partial_tS)}{t^2},
\eea
where $C_2, C_3$ are additional integration constants. We stress that this metric is the {\it most general} solution of the reverse-engineering procedure, given the confines of the initial assumption that $Y_{\frac{1}{2}}$ becomes a bulk Killing vector while remaining unchanged.
However, we cannot claim that we have identified the holographic dual of a general theory possessing the full symmetry of the Age algebra. The reason for this is that we are able to identify one more Killing vector of the metric compatible with the following bulk extension of $X_{-1}$
\bea
&&\!\!\!\!\!\!-X_{-1}=\partial_t  
 - \frac{\partial_t g_{rr}}{\partial_z g_{rr}}\partial_z\nn\\
&&\!\!\!\!\!\!
+i\bigg(-\frac{2 C_2}{t^2 g_{rr}}+ \frac{S}{2t^2} - \frac{\partial_t S+2T +4C_1}{2t} + \frac{\partial_t g_{rr}\partial_z S}{2t \partial_z g_{rr}}
\bigg)\partial_\xi .\nn\\
\eea 
Thus the isometries of the above metric generate the full Schr\"{o}dinger algebra, as in \cite{Jottar:2010vp}. Naturally the correlators computed from this metric using holography exhibit the kind of "fake" aging discussed earlier, and are constrained by the full Schr\"odinger algebra.
For the time being we can only trace this feature to the assumption made regarding the bulk realization of $Y_{\frac{1}{2}}$. (This assumption was also made in \cite{Jottar:2010vp}.) Relaxing this condition makes the problem of identifying the holographic metric of Aging much more complicated. We leave this question for our future work.

In conclusion, in this letter we have clearly pointed out the difference between the aging dynamics realized as
``dressed'' Schr\"odinger dynamics from pure aging. In particular we have obtained the 3-point functions for aging which {\it cannot} be obtained by ``dressing'' the 3-point Schr\"odinger correlators.  The physical implications of our new results are yet to be understood.
Nevertheless, it is reasonable to expect that these results will have practical importance in
the real and numerical experiments of aging dynamics \cite{agingbook} and that they should be generalizable to the relativistic context, with possible applications to the physics of the quark-gluon plasma. We plan to address these issues in our future work.


{\bf Acknowledgements:}
We thank Rob Leigh, Leo Pando Zayas and Michel Pleimling for insightful discussions and, 
in particular, we thank Malte Henkel for generous and incisive comments on this work. {DM}
is supported, in part, by the U.S. Department of Energy
under Grant No. DE-FG05-92ER40677, and 
{DV and CW} under Grant No. DEFG02-
97ER41027.

\end{document}